\documentstyle[12pt]{article}
\renewcommand{\O}{\Omega}
\renewcommand{\o}{\theta}
\newcommand{\g}{{\scriptstyle \cal G\mit}}
\renewcommand{\b}{{\scriptstyle \cal B\mit}}
\newcommand{\n}{{\scriptstyle \cal N\mit}}
\newcommand{\h}{{\scriptstyle \cal H\mit}}
\renewcommand{\P}{{\cal P}}
\renewcommand{\d}{{\scriptstyle \cal D\mit}}
\newcommand{\m}{\mu}
\newcommand{\e}{\varepsilon}
\newtheorem{fact}{Theorem}

\begin{document}
\begin{titlepage}
\title{\begin{flushright} {\normalsize PAR-LPTHE 93-08 \\
UUITP 5/1993   \\
hep-th/9303038 } \\
\end{flushright}  \bigskip   \bigskip
{\bf Symplectic structures
\\associated to Lie-Poisson groups.}}
\author{\\ {\bf A. Yu. Alekseev}
\thanks{On leave of absence from
LOMI, Fontanka 27, St.Petersburg, Russia.}
\\\\
Laboratoire de Physique Th\'{e}orique et Hautes
\'{E}nergies\thanks{Unit\'{e} Associ\'{e}e au C.N.R.S.,
URA 280.}\ ,\\ Paris, France\thanks{LPTHE, Paris-VI,
Tour 16 - 1er \'{e}tage, 4 place Jussieu, F-75252
PARIS CEDEX 05.}\ .\\\\ {\bf A. Z. Malkin}
\thanks{Supported
in part by a Soros Foundation Grant awarded by
the American Physical Society.}\ \thanks{On leave of
absence from St.Petersburg University.}\\\\
Institute of Theoretical Physics, Uppsala University,
\\ Box 803 S-75108, Uppsala, Sweden.\\\\}
\date{23 February 1993}
\maketitle \thispagestyle{empty}
\begin{abstract}
The Lie-Poisson analogues of the cotangent bundle
and coadjoint orbits of a Lie
group are considered. For the natural Poisson brackets
the symplectic leaves in these manifolds are classified
and the corresponding symplectic forms
are described.  Thus the construction of the Kirillov
symplectic form is generalized for Lie-Poisson groups.
\\\\\\
\end{abstract}
\end{titlepage}
\section*{Introduction.}

The method of geometric quantization \cite{1} provides
a set of Poisson manifolds associated to each Lie
group $G$. The dual space $\g^{*}$ of the corresponding
Lie algebra $\g$ plays an important role in this theory.
The space $\g^{*}$ carries the \mbox{Kirillov-Kostant}
Poisson bracket which mimics the Lie commutator in $\g$.
Having chosen a basis $\{\e^{a}\}$ in $\g$, we can
define structure constants $f^{ab}_{c}\,$:
\begin{equation}
[\e^{a},\e^{b}\:]=\sum_{c} f^{ab}_{c}\e^{c} \ \
,  \end{equation}
where [,]\, is the Lie commutator in $\g$.
216z
On the other hand, we can treat any element $\e^{a}$
of the basis as a linear function on $\g^{*}$.
The Kirillov-Kostant Poisson bracket is defined so that
it resembles formula (1):
\begin{equation} \{\e^{a},\e^{b}\:\}=\sum_{c}
f^{ab}_{c}\e^{c} \ \ ,\end{equation}

The Kirillov-Kostant bracket has two important
properties :
\begin{enumerate}
\item the r.h.s. of (2) is linear in $\e^{c}$,
\item the group $G$\, acts on $\g^{*}$ by means of
the coadjoint action and preserves the bracket (2).
\end{enumerate}

The Kirillov-Kostant bracket is always degenerate
(\mbox{e.\ g.\ } at the origin in  $\g^{*}$).
According to the general theory of Poisson
manifolds \cite{2,3} the space $\g^{*}$ splits into
the set of symplectic leaves. Usually it is not easy
to describe symplectic leaves of a Poisson manifold.
Fortunately an effective description exists in this
very case. Symplectic leaves coincide with orbits of
the coadjoint action of $G\,$ in $\g^{*}$.
Kirillov obtained an elegant expression for the
symplectic form $\O$ on the orbit \cite{1}:
\begin{equation} \O_{X}(u,v)=<X,[\e_{u},\e_{v}]>   \ \
.\end{equation}
Here $<\,,>$ is the canonical pairing between $\g$
and $\g^{*}$. The value of the form is calculated
at the point $X$ on the pair of vector fields $u$
and $v$ on the orbit. The elements $\e_{u}, \e_{v}$ of
the algebra $\g$\, are defined as follows:
\begin{equation}  u\!\!\mid_{_{X}}=ad^{*}(\e_{u})X  \ \
,    \end{equation}
where $\:ad^{*}$ is the coadjoint action of $\g$
on $\g^{*}$. The purpose of this paper is to generalize
formula (3) for  Lie-Poisson groups.

Lie group $G$ equipped with a Poisson bracket \{,\}
is called a Lie-Poisson group when the
multiplication in $G$
\begin{equation} G\times G\longrightarrow
G\end{equation}
\begin{equation} (g,g^{'})\longrightarrow
gg^{'} \end{equation}
is a Poisson mapping. In other words, the bracket
of any two functions $f$ and $h$ satisfies the
following condition:
\begin{equation} \{f,h\}(gg^{'})=\{f(gg^{'})
,h(gg^{'})\}_{g}+\{f(gg^{'}),h(gg^{'})\}_{g^{'}}  \ \ .
\end{equation}
Here we treat $f(gg^{'}),\, h(gg^{'})$ as functions
of the argument $g$ only in the first term of
the r.h.s. whereas in the second term they
are considered as functions of $g^{'}$.

In the framework of the Poisson theory the natural
action of a group on a manifold is the Poisson
action \cite{4,5}. It means that the mapping
\begin{equation}  G\times M\longrightarrow
M  \end{equation}
is a Poisson one. In Poisson theory this property
replaces property (ii) of Kirillov-Kostant bracket.
There exist direct analogues of the coadjoint
orbits for Lie-Poisson groups. Our goal in
this paper is to obtain an analogue of formula (3).
However, it is better to begin with Lie-Poisson
analogue of the cotangent bundle $T^{*}\! G$
described in section 2. The symplectic form for
this case is obtained in section 3 and then in
section 4 the analogue of the Kirillov form
appears as a result of reduction.
Section 1 is devoted to an exposition of the
Kirillov theory. In section 5 some examples
are considered.

When speaking about Lie-Poisson theory
the works of Drinfeld \,\cite{6}
, \,Semenov-Tian-Shansky \cite{5}
, \, Weinstein and Lu \cite{7} must be mentioned.
We follow these papers when representing
the known results.

The theory of Lie-Poisson groups is
a quasiclassical version of the theory of
quantum groups. So we often use
the attribute ``deformed'' instead of ``Lie-Poisson''.
Similarly we call the case when the Poisson bracket
on the group is equal to zero the ``classical'' one.

\section{Symplectic structures associated
to Lie groups.}

For the purpose of selfconsistency
we shall collect in this section some
well-known results concerning Poisson
and symplectic geometry associated to Lie groups.
The most important part of our brief survey is
a theory of coadjoint orbits. Our goal is
to rewrite the Kirillov symplectic form so that
a generalization can be made straightforward.

Let us fix notations. The main object
of our interest is a Lie group $G$.
We denote the corresponding Lie algebra by $\g$.
The linear space $\g$ is supplied with
Lie commutator [,]. If $\{\e^{a}\}$ is a basis
in $\g$ we can define structure
constants $f^{ab}_{c}$ in the following way:
\begin{equation}
[\e^{a},\e^{b}\:]=\sum_{c} f^{ab}_{c}\e^{c}  \ \
.\end{equation}

The Lie group $G$ has a representation
which acts in $\g$. It is called adjoint
representation:
\begin{equation}  \e^{g}\equiv Ad (g)\e    \ \
.\end{equation}
The corresponding representation of the
algebra $\g$ is realized by the commutator:
\begin{equation}  ad (\e )\eta =[\e ,\eta ] \ \
.   \end{equation}
We denote elements of the algebra $\g$ by
small Greek letters.

Let us introduce a space $\g^{*}$ dual to
the Lie algebra $\g$. There is a canonical
pairing $<\,,>$ between $\g^{*}$ and $\g$
and we may construct a basis $\{l_{a}\}$
in $\g^{*}$ dual to the basis $\{\e^{a}\}$
so that
\begin{equation}   <l_{a}\,
,\e^{b}>=\delta _{a}^{b} \ \ .\end{equation}
We use small Latin letters for elements
of $\g^{*}$. Each vector $\e$ from $\g$
defines a linear function on $\g^{*}$:
\begin{equation}  H_{\e}(l)\! =<l,\e>  \ \
.    \end{equation}
In particular, a linear function $H^{a}$
corresponds to an element $\e^{a}$ of the basis in $\g$.

By duality the group $G$ and its Lie algebra $\g$
act in the space $\g^{*}$ via
the coadjoint representation:
\begin{equation} <Ad ^{*}(g)l,\e>=<l,Ad(g^{-1})\e >
\ \ ,  \end{equation}
\begin{equation} <ad ^{*}(\e)l,\eta >=-<l,[\e,\eta ]>
\ \ .  \end{equation}

The space $\g$ can be considered as a space
of left-invariant or right-invariant vector fields
on the group $G$. Let us define the universal
right-invariant one-form $\o _{g}$ on $G$
which takes values in $\g \,$:
\begin{equation}  \o _{g}(\e)=-\e \ \ . \end{equation}
We treat $\e$ in the l.h.s. of formula (16) as
a right-invariant vector field whereas in the r.h.s.
as an element of $\g$. Since the one-form $\o _{g}$
and  the vector field $\e$ are right-invariant
the result does not depend on the point $g$ of
the group. $\o_{g}$ is known as Maurer-Cartan form.

Similarly, the universal left-invariant
one-form $\m_{g}$ can be introduced:
\begin{equation} \m_{g}(\e )=\e \ \
,\ \  \m_{g}=Ad(g^{-1})\o_{g}  \ \ ,  \end{equation}
where $\e$ is a left-invariant vector field,
$Ad$ acts on values of $\o_{g}$.

In the case of matrix group $G$ the invariant
forms $\o_{g}$ and $\m_{g}$ look like follows:
\begin{equation} \o_{g}=dg\,g^{-1}  \ \
,   \end{equation}
\begin{equation} \m_{g}=g^{-1}dg  \ \
.   \end{equation}

For any group $G$ there exist
two covariant differential operators
$\nabla \! \! _{L}$ and $\nabla \! \! _{R}$
taking values in the space $\g^{*}$.
These  are left and right derivatives:
\begin{equation} <\nabla \! \!  _{L}f,\e >(g)
=-\frac{d}{dt} f(exp(t\e)g)  \ \
,       \end{equation}  \begin{equation}
<\nabla \! \!  _{R}f,\e>(g)=\frac{d}{dt}
f(g\, exp(t\e))  \ \ . \end{equation}
where $exp$ is the exponential map from
a Lie algebra to a Lie group. The simple relation
for left and right derivatives
of the same function $f$ holds:
\begin{equation} \nabla \! \!  _{R}f=-Ad^{*}(g^{-1})\nabla
\! \!  _{L}f \ \ . \end{equation}

{}From the very beginning the linear space $\g^{*}$ is not
supplied with a natural commutator.
Nevertheless, we define
the commutator $[,\!]^{*}$ in $\g^{*}$
and put it equal to zero:
\begin{equation} [l,m]^{*}=0  \ \
.  \end{equation}
The main technical difference of
the deformed theory from the classical one is
that the commutator in $\g^{*}$ is nontrivial.
As a consequence, the corresponding group $G^{*}$
becomes nonabelian. This fact plays a crucial role
in the consideration of Lie-Poisson theory.
In the classical case the Lie algebra $\g^{*}$
is just abelian and the group $G^{*}$ coincides
with $\g^{*}$.

The space $\g^{*}$ carries a natural Poisson structure
invariant with respect to the coadjoint action
of $G$ on $\g^{*}$. Let us remark that the differential
of any function on $\g^{*}$ is an element of the dual space
, i.e. of the Lie algebra $\g$. It gives us a possibility
to define the following Kirillov-Kostant Poisson bracket:
\begin{equation} \{f,h\}(l)=<l,[df(l),dh(l)]>  \ \
.     \end{equation}
In particular, for linear functions $H_{\e}$ the r.h.s.
of (24) simplifies:
\begin{equation} \{H_{\e},H_{\eta}\}=H_{[\e,\eta]}  \ \
,  \end{equation}
\begin{equation} \{H^{a},H^{b}\}
=\sum_{c}f^{ab}_{c}H^{c} \ \ .  \end{equation}
The last formula simulates the commutation relations (1).

In general situation the space $\g^{*}$ supplied
with Poisson bracket (24) is not a symplectic manifold.
The Kirillov-Kostant bracket is degenerate.
For example, in the simplest case of $\g=su(2)$
the space $\g^{*}$ is 3-dimensional.
The matrix of Poisson bracket is antisymmetric
and degenerates as any antisymmetric matrix
in an odd-dimensional space.

The relation between symplectic and Poisson theories
is the following. Any Poisson manifold
with degenerate Poisson bracket splits into a set
of symplectic leaves. A symplectic leaf is defined
so that its tangent space at any point consists
of the values of all hamiltonian vector fields
at this point:
\begin{equation}  v_{h}(f)=\{h,f\} \ \
.  \end{equation}
Each symplectic leaf inherits the Poisson bracket from
the manifold. However, being restricted
onto the symplectic leaf the Poisson bracket becomesnondegenerate and we can
define the symplectic
two-form $\O$ so that:
\begin{equation} \O (v_{f},v_{h})=\{f,h\} \ \
.   \end{equation}
The relation (28) defines $\O$ completely
because any tangent vector to the symplectic
leaf may be represented as a value
of some hamiltonian vector field.

If we choose dual bases $\{e_{a}\}$ and $\{e^{a}\}$ in
tangent and cotangent spaces to the symplectic leaf
we can rewrite the bracket and the symplectic form
as follows:
\begin{equation}  \{f,h\}=-\sum
_{ab}P^{ab}<df,e_{a}><dh,e_{b}>  \ \ , \end{equation}
\begin{equation}  \O=\sum _{ab}\O_{ab}\, e^{a}\!\!
 \otimes \! e^{b}=
\frac{1}{2}\sum _{ab}\O_{ab} \, e^{a}\!\! \wedge \!
\e^{b} \ \ . \end{equation}
Using definition (28) of the form $\O$ and
formulae (29),(30) one can check that
the matrix $\O_{ab}$ is inverse to the matrix $P^{ab}$:
\begin{equation}  \sum _{c}\O_{ac}P^{cb}=\delta _{a}^{b}
\ \  .  \end{equation}

For the particular case of the space $\g^{*}$ with
Poisson structure (24), there exists
a nice description of the symplectic leaves.
They coincide with the orbits of \mbox{coadjoint}
action (14) of the group $G$.
Starting from any point $l_{0}$,
we can construct an orbit
\begin{equation} O_{l_{0}}=\{l=Ad^{*}(g)l_{0}\ \
,\ \ g\in G\} \ \  . \end{equation}
Any point of $\g^{*}$ belongs to some coadjoint orbit.
The orbit $O_{l_{0}}$ can be regarded as a  quotient
space of the group $G$ over its subgroup $S_{l_{0}}$:
\begin{equation}  O_{l_{0}}\approx G/S_{l_{0}}  \ \
,  \end{equation}
where $S_{l_{0}}$ is defined as follows:
\begin{equation} S_{l_{0}}=\{g\in G\ \
,\ \ Ad^{*}(g)l_{0}=l_{0}\} \ \  .\end{equation}

In the case of $G=SU(2)$ the coadjoint action
is represented by rotations in the 3-dimensional
space $\g^{*}$. The orbits are spheres
and there is one exceptional zero radius orbit
which is just the origin.
The group $S_{l_{0}}$ is isomorphic to $U(1)$
and corresponds to rotations around
the axis parallel to $l_{0}$. For the exceptional
orbit $S_{l_{0}}=G$ and the quotient space $G/G$ is a point.

Let us denote by $p_{l_{0}}$ the projection from
$G$ to $O_{l_{0}}$:
\begin{equation} p_{l_{0}}:\;\;\;g\longrightarrow
l_{g}=Ad^{*}(g)l_{0}  \ \  .
\end{equation}
We may investigate the symplectic form $\O$ on
the orbit directly. However, for technical reasons
it is more convenient to consider its pull-back
$\O_{l_{0}}^{G}=p^{*}_{l_{0}}\O$ defined
on the group $G$ itself.
We reformulate the famous Kirillov's result
in the following form.
Let $O_{l_{0}}$ be a coadjoint orbit of the group $G$
and $p_{l_{0}}$ be  the projection (35).
The Poisson structure (24) defines
a symplectic form $\O$ on      $\! \! O_{l_{0}}$.
\begin{fact}
The pull-back of $\O$ along the projection $p_{l_{0}}$
is the following:
\begin{equation} \O_{l_{0}}^{G}=\frac{1}{2}
<dl_{g}\stackrel{\wedge}{,}\o_{g}>   \ \ .\end{equation}
\end{fact}

We do not prove formula (36) but the proof of
its Lie-Poisson counterpart in section 3
will fill this gap. Let us make only few  remarks.
First of all, the form $\O_{l_{0}}^{G}$ actually is
a pull-back of some two-form on the orbit $O_{l_{0}}$.
Then, $\O_{l_{0}}^{G}$ is a closed form:
\begin{equation} d\O_{l_{0}}^{G}=0  \ \  .\end{equation}
This is a direct consequence of the Jacobi identity for
the Poisson bracket (24). The form $\O_{l_{0}}^{G}$
is exact, while the original form $\O$ belongs to
a nontrivial cohomology class.
The left-invariant one-form
\begin{equation} \alpha =<l_{g},\o_{g}>=<l_{0},\m_{g} >
\end{equation}
satisfies the equation
\begin{equation}   d\alpha=\O_{l_{0}}^{G} \ \  .
\end{equation}

In physical applications the form $\alpha$ defines
an action for a hamiltonian system on the orbit:
\begin{equation} S=\int \alpha   \ \ .  \end{equation}

Returning to the formula (36) we shall speculate
with the definition of $G^{*}$. In our case $G^{*}=\g^{*}$
and we may treat $l_{g}$ as an element of $G^{*}$.
For an abelian group Maurer-Cartan forms $\o$
and $\m$ coincide with the differential
of the group element:
\begin{equation}  \o_{l}=\m_{l}=dl \ \ .  \end{equation}
Using (41) we rewrite (36):
\begin{equation} \O_{l_{0}}^{G}=\frac{1}{2}<\o_{l}
\stackrel{\wedge}{,}\o_{g}>   \ \ ,\end{equation}
where $l$ is the function of $g$ given by formula (35).
Expression (42) admits a straightforward generalization
for Lie-Poisson case.

The rest of this section is devoted
to the cotangent bundle $T^{*}\!G$
of the group $G$. Actually, the bundle $T^{*}\!G$
is trivial. The group $G$ acts on itself
by means of right and left multiplications.
Both these actions may be used to trivialize $T^{*}\!G$.
So we have two parametrizations of
\begin{equation}  T^{*}\!G=G\times \g^{*}  \end{equation}
by pairs $(g,l)$ and $(g,m)$ where $l$ and $m$ are
elements of $g^{*}$. In the left parametrization $G$
acts on $T^{*}\!G$ as follows:
\begin{equation} \rm L\ \ \ \ \ \ \ \ \ \ \mit
h:\; (g,m)\longrightarrow (hg,m) \ \ , \ \ \ \ \
\ \ \ \end{equation}
\begin{equation} \rm R\ \ \ \ \ \mit h:\;
(g,m)\longrightarrow (gh^{-1},Ad^{*}(h)m) \ \
.\end{equation}
In the right parametrization left and
right multiplications change roles:
\begin{equation} \rm L\ \ \ \ \ \ \mit h:\;
(g,l)\longrightarrow (hg,Ad^{*}(h)l) \ \
,\end{equation}
\begin{equation} \rm R\ \ \ \ \ \ \ \ \ \
\mit h:\;(g,l)\longrightarrow (gh^{-1},l) \ \
.\ \ \end{equation}
The two coordinates $l$ and $m$ are related:
\begin{equation}  l=Ad^{*}(g)m  \ \ .  \end{equation}

The cotangent bundle $T^{*}\!G$ carries
the canonical symplectic structure $\O^{T^{*}\!G}$
\cite{2}. Using coordinates $(g,l,m)$, we write
a formula for $\O^{T^{*}\!G}$ without the proof:
\begin{equation} \O^{T^{*}\!G}=\frac{1}{2}
(<dm\stackrel{\wedge}{,}\m_{g}>
+<dl\stackrel{\wedge}{,}\o_{g}>) \ \ . \end{equation}

The symplectic structure on $T^{*}\!G$
is a sort of universal one.
We can recover the Kirillov two-form (36)
for any orbit starting from (49). More exactly,
let us impose in (49) the condition:
\begin{equation} m=m_{0}=const \ \  .  \end{equation}
It means that instead of $T^{*}\!G$ we consider
a reduced symplectic manifold with
the symplectic structure
\begin{equation}  \O_{r}=\frac{1}{2}<dl,\o_{g}> \ \
,  \end{equation}
where $l$ is subject to constraint
\begin{equation}  l=Ad^{*}(g)m_{0} \ \
.  \end{equation}
Formulae (51), (52) reproduce formulae (35), (36)
and we can conclude that the reduction leads
to the orbit $O_{m_{0}}$ of the point $m_{0}$
in $\g^{*}$.

The aim of this paper is to present
Lie-Poisson analogues of formulae (36) and (49).
Having finished  our sketch of the classical theory,
we pass to the deformed case.

\section{Heisenberg double of Lie bialgebra.}

One of the ways to introduce deformation
leading to Lie-Poisson groups is to consider
the bialgebra structure on $\g$. Following \cite{6},
we consider a pair $(\g ,\g^{*})$, where we treat $\g^{*}$
as another Lie algebra with the commutator $[,\!]^{*}$.
For a given commutator $[,\!]$ in $\g$ we can not choose
an arbitrary commutator $[,\!]^{*}$ in $\g^{*}$.
The axioms of bialgebra can be reformulated as follows.
The linear space
\begin{equation}    \d =\g +\g^{*}   \end{equation}
with the commutator $[,\!]_{\d}$:
\begin{equation}  [\e,\eta]_{\d}=[\e,\eta] \ \
,   \end{equation}
\begin{equation}  [x,y]_{\d}=[x,y]^{*}  \ \ ,
\end{equation}
\begin{equation}  [\e,x]_{\d}=ad^{*}(\e )x
-ad^{*}(x)\e  \ \   .  \end{equation}
must be a Lie algebra.
In the last formula (56) $ad^{*}(\e )$ is
the usual $ad^{*}$-operator for the Lie algebra $\g$
acting on $\g^{*}$. The symbol $ad^{*}(x)$
corresponds to the coadjoint action
of the Lie algebra $\g^{*}$ on its dual space $\g$.

The only thing we have to check is
the Jacobi identity for the commutator $[,\!]_{\d}$.
If it is satisfied, we call the pair $(\g ,\g^{*})$
Lie bialgebra. Algebra $\d$ is called Drinfeld double.
It has the nondegenerate scalar product $<\, ,>_{\d}$ :
\begin{equation} <(\e,x),(\eta,y)>_{\d}=<y,\e >
+<x,\eta >  \ \ ,  \end{equation}
where in the r.h.s. $<\, ,>$ is the canonical pairing
of $\g$ and $\g^{*}$. It is easy to see that
\begin{equation} < \g,\g >_{\d}=0\ \
,\ \ <\g^{*},\g^{*}>_{\d}=0 \ \  . \end{equation}
In other words, $\g$ and $\g^{*}$ are isotropic
subspaces in $\d$ with respect to the form $<\, ,>_{\d}$.
We call the form $<\, ,>_{\d}$ on the algebra $\d$
standard product in $\d$.

We shall need two operators $P$ and $P^{*}$
acting in $\d$. $P$ is defined as
a projector onto the subspace $\g$:
\begin{equation}  P(x+\e)=\e \ \
.  \end{equation}

The operator $P^{*}$ is its conjugate
with respect to form (57). It  appears to be
a projector onto the subspace $\g^{*}$:
\begin{equation}  P^{*}(x+\e)=x  \ \
.  \end{equation}

The standard product in $\d$ enables us
to define the canonical isomorphism $J\ :\ \
\d^{*}\longrightarrow \d$ by means of the formula
\begin{equation} <J(a^{*}),b>_{\d}=<a^{*},b> \ \
, \end{equation}
where $a^{*}$ is an element of $\d^{*}$ and $b$
belongs to $\d$. In the r.h.s. we use
the canonical pairing of $\d$ and $\d^{*}$.
The standard product can be defined
on the space $\d^{*}$:
\begin{equation}
<a^{*},b^{*}>_{\d^{*}}=<J(a^{*}),J(b^{*})>_{\d}  \ \
,  \end{equation}
where $a^{*}$ and $b^{*}$ belong to $\d^{*}$.
The scalar product $<\, ,>_{\d}$ is invariant
with respect to the commutator in $\d$:
\begin{equation}  <[a,b],c>_{\d}+<b,[a,c]>_{\d}=0  \ \
.  \end{equation}
It is easy to check that the operator $J$
converts $ad^{*}$ into $ad$:
\begin{equation}  Jad^{*}(a)J^{-1}=ad(a) \ \
. \end{equation}

Using the standard scalar product in $\d$,
one can construct elements $r$ and $r^{*}$ in $\d
\otimes \d$ which correspond to the operators $P$
and $P^{*}$:
\begin{equation}
<a\!\otimes\! b,r>_{\d\otimes\d}=<a,Pb>_{\d}  \ \
, \end{equation}
\begin{equation}
<a\!\otimes\! b,r^{*}>_{\d\otimes\d}=-<a,P^{*}b>_{\d}
\ \ .
\end{equation}
In terms of dual bases $\{ \e^{a}\}$
and $\{ l_{a}\}$ in $\g$ and $\g^{*}$
\begin{equation}  r=\sum_{a}\e^{a}\otimes
l_{a} \ \ \ ,\ \ \ r^{*}=-\sum_{a}l_{a}\otimes
\e^{a} \ \  .  \end{equation}

The Lie algebra $\d$ may be used to construct
the Lie group $D$. We suppose that $D$
exists (for example, for finite dimensional algebras
it is granted by the Lie theorem) and we choose it
to be connected. Originally the double is defined as
a connected and simply connected group.
However, we may use any connected group D
corresponding to Lie algebra $\d$.
Property (64) can be generalized for $Ad$
and $Ad^{*}$:
\begin{equation}  JAd^{*}(d)J^{-1}=Ad(d) \ \
, \end{equation}
where $d$ is an element of $D$.

Let us denote by $G$ and $G^{*}$ the subgroups
in $D$ corresponding to subalgebras $\g$ and $\g^{*}$
in $\d$. In the vicinity of the unit element of $D$
the following two decompositions are applicable:
\begin{equation}  d=gg^{*}=h^{*}h  \ \  ,  \end{equation}
where $d$ is an element of $D$, coordinates $g,h$
belong to the subgroup $G$, coordinates $g^{*},h^{*}$
belong to the subgroup $G^{*}$.

To generalize formula (69), let us consider
the set $\Im$ of classes $G\!\!\setminus
\!\! D\! /\! G^{*}$. We denote individual classes
by small letters $i,j,\ldots$.
Let us pick up a representative $d_{i}$
in each class $i$. If an element $d$ belongs
to the class $i$, it can be represented in the form
\begin{equation}   d=gd_{i}g^{*} \end{equation}
for some $g$ and $g^{*}$. In general case
the elements $g$ and $g^{*}$ in  decomposition  (70)
are not defined uniquely. If $S(d_{i})$ is
a subgroup in $G$
\begin{equation}  S(d_{i})=\{ h\in G\ \
,\ \ d_{i}^{-1}hd_{i}\in G^{*}\}
\ \ , \end{equation}
 we can take a pair $(gh,d_{i}^{-1}h^{-1}d_{i}g^{*})$
instead of $(g,g^{*})$, where $h$ is
an arbitrary element of $S(d_{i})$.
We denote $T(d_{i})$ the corresponding
subgroup in $G^{*}$:
\begin{equation} T(d_{i})=d_{i}^{-1}S(d_{i})d_{i}
\ \  . \end{equation}
So we have the following stratification
of the double $D$:
\begin{equation}
D=\bigcup_{i\in \Im}Gd_{i}G^{*}=\bigcup_{i\in \Im}C_{i}
\ \  . \end{equation}
Each cell
\begin{equation}  C_{i}=Gd_{i}G^{*}   \end{equation}
in this decomposition is isomorphic to the quotient
of the direct product $G\times G^{*}$ over $S(d_{i})$,
where
\begin{equation}  (g,g^{*})\sim (g^{'},g^{*'}) \ \ \
\ \ {\rm if} \end{equation} \begin{equation}
g^{'}=gh\ \ ,\ \ g^{*'}=d_{i}^{-1}h^{-1}d_{i}g^{*}\ \
,\ \ h\in S(d_{i}) \ \  . \end{equation}

For the inverse element $d^{-1}$
in the relation (70) we get another stratification
of $D$ in which $G$ and $G^{*}$ replace each other:
\begin{equation} D=\bigcup_{i\in \Im
}G^{*}d_{i}^{-1}G=\bigcup_{i\in \Im}c_{i}
\ \ . \end{equation}

Now we turn to the description of the Poisson brackets
on the manifold $D$. Double $D$ admits two natural
Poisson structures. First of them was proposed by
Drinfeld \cite{6}. For two functions $f$
and $h$ on $D$ the Drinfeld bracket is equal to
\begin{equation} \{ f,h\} =<\nabla \! \! _{L}f\otimes
\nabla \! \! _{L}h,r>-
<\nabla \! \! _{R}f\otimes \nabla \! \! _{R}h,r> \ \
,   \end{equation}
where $<\, ,>$ is the canonical pairing between $\d
\otimes \d$ and $\d^{*} \otimes \d^{*}$.
Poisson bracket (78) defines a structure of
a Lie-Poisson group on $D$. However, the most important
for us is the second Poisson structure on $D$
suggested by Semenov-Tian-Shansky \cite{5}:
\begin{equation} \{ f,h\} =-(<\nabla \! \! _{L}f\otimes
\nabla \! \! _{L}h,r>+
<\nabla \! \! _{R}f\otimes \nabla \! \! _{R}h,r^{*}>)
\ \  .  \end{equation}

The manifold $D$ equipped with bracket (79) is called
Heisenberg double or $D_{+}$.
It is a natural analogue of $T^{*}\!G$
in the Lie-Poisson case.
When $\g^{*}$ is abelian, $G^{*}=\g^{*}$
and $D_{+}=T^{*}\!G$. If the double $D$ is
a matrix group, we can rewrite
the basic formula (79) in the following form:
\begin{equation} \{ d^{1},d^{2}\} =-(rd^{1}d^{2}
+d^{1}d^{2}r^{*})  \ \ ,  \end{equation}
where $d^{1}=d\otimes I\ ,\ d^{2}=I\otimes d$.

The problem which appears immediately
in the theory of $D_{+}$ is the possible degeneracy
of Poisson structure (79) in some points of $D$.
It is important  to describe the stratification
of $D_{+}$ into the set of symplectic leaves.
The answer is given by the following
\begin{fact}
Symplectic leaves of $D_{+}$ are connected components
of nonempty intersections of left
and right stratification cells:
\begin{equation}
D_{ij}=C_{i}\cap c_{j}=Gd_{i}G^{*}\cap
G^{*}d_{j}^{-1}G \ \ . \end{equation}
\end{fact}
{\em Proof.} The tangent space $T_{d}^{S}$
to the symplectic leaf at the point $d$ coincides with
the space of values of all hamiltonian vector fields
at this point. For concrete calculations
let us choose the left identification
of the tangent space to $D$ with $\d$.
We can rewrite the Poisson bracket (79) in terms
of left derivatives $\nabla \! \! _{L}$:
\begin{eqnarray} \{ f,h\} (d) =-(<\nabla \!
\!_{L}f\otimes \nabla \!
\! _{L}h,r>+<Ad^{*}(d^{-1})\nabla \! \! _{L} f\otimes
Ad^{*}(d^{-1})\nabla \! \! _{L}h,r^{*}>)=   \nonumber
\\
=-<\nabla \! \! _{L}f\otimes \nabla \! \! _{L}h,r
+Ad(d)\otimes Ad(d)\, r^{*}> \ \  . \ \ \ \ \ \ \ \
\ \ \ \  \end{eqnarray}
Here we use relation (22) between left
and right derivatives on a group.

A hamiltonian $h$ produces the hamiltonian
vector field $v_{h}$ so that the formula
\begin{equation} <df,v_{h}>=\{ h,f\}  \end{equation}
holds for any function $f$.
Using (82), (83) we can reconstruct the field $v_{h}$:
\begin{equation}  v_{h}=<\nabla \! \! _{L}h
,r+Ad(d)\otimes Ad(d)\, r^{*}>_{2}  \ \ .\end{equation}
Having identified $\d$ and $\d^{*}$ by means
of the operator $J$, we can rewrite the r.h.s. of (84)
as follows:
\begin{equation}  v_{h}|_{_{{\displaystyle d}}}=\P
dh=(P-Ad(d)P^{*}Ad(d^{-1}))J(\nabla \! \! _{L}h(d))
\ \ , \end{equation}
where $\P$ acts in $\d$:
\begin{equation} \P=P-Ad(d)P^{*}Ad(d^{-1})  \ \
.  \end{equation}
It is called Poisson operator.
Using the fact that the value of $\nabla \! \! _{L}h$
at the point $d$ is an arbitrary vector from $\d^{*}$,
we conclude that $T_{d}^{S}$ coincides
with the image of the operator $\P$:
\begin{equation}  T_{d}^{S}=Im\P  \ \  .\end{equation}
The most simple way to describe the image of $\P$
is to use the property:
\begin{equation}  Im\P =(Ker\P^{*})^{\perp}  \ \
. \end{equation}
Here conjugation and symbol $\perp$ correspond
to the standard product in $\d$.
The operator $\P^{*}$ is given by the formula
\begin{equation}  \P^{*}=P^{*}-Ad(d)PAd(d^{-1})
\ \   .  \end{equation}

Suppose that a vector $a=x+\e$ belongs to $Ker\P^{*}$:
\begin{equation} \P^{*}(x+\e)=0  \ \  .  \end{equation}
Let us rewrite the condition (90) in the following form:
\begin{equation} (Ad(d^{-1})P^{*}-P Ad(d^{-1}))(x+\e)=0
\ \  ,  \end{equation}
or, equivalently,
\begin{equation}  Ad(d^{-1})x=P (Ad(d^{-1})x
+Ad(d^{-1})\e) \ \  . \end{equation}
Using the property
\begin{equation}  P+P^{*}=id \end{equation}
of the projectors $P$ and $P^{*}$,
one can get from (92):
\begin{equation}   P^{*}(Ad(d^{-1})x)=P(Ad(d^{-1})\e)
\ \  . \end{equation}
The l.h.s. of (94) is a vector from $\g^{*}$
whereas the r.h.s. belongs to $\g$. So the equation (94)
implies that both the l.h.s. and the r.h.s.
are equal to zero.

Let $V(d)$ be the subspace in $\g$ defined
by the following condition:
\begin{equation}  V(d)=\{ \e\in \g\ \
,\ \ Ad(d^{-1})\e \in \g^{*}\} \ \  .   \end{equation}
In the same way we define the subspace $V^{*}(d)$
in $\g^{*}$:
\begin{equation}   V^{*}(d)=\{x\in \g^{*}\ \
,\ \ Ad(d^{-1})x\in\g \} \ \  .   \end{equation}
It is not difficult to check that $V(d)$
and $V^{*}(d)$ are actually Lie subalgebras in $\g$
and $\g^{*}$. The kernel of the operator $\P^{*}$
may be represented as a direct sum of $V(d)$
and $V^{*}(d)$:
\begin{equation}   Ker\P^{*}=V(d)\oplus V^{*}(d) \ \
.  \end{equation}
The tangent space $T^{S}_{d}$ to the symplectic leaf
at the point $d$ acquires the form
\begin{equation}  T^{S}_{d}=(V(d)\oplus
V^{*}(d))^{\perp} \ \   . \end{equation}
The result (98) can be rewritten:
\begin{equation}   T_{d}^{S}=V(d)^{\perp}\cap
V^{*}(d)^{\perp}=(V(d)^{\perp}\cap \g^{*})\oplus
(V^{*}(d)^{\perp}\cap \g ) \ \   . \end{equation}
Here the last expression represents $T^{S}_{d}$
as a direct sum of its intersections with $\g$
and $\g^{*}$.

Now we must compare subspace (99)
with the tangent space $T_{d}^{'}$
of the intersection of the stratification
cells (theorem 2).
Suppose that the point $d$ belongs
to the cell $D_{ij}$ of the stratification.
We can rewrite the definition of $D_{ij}$
as follows:
\begin{equation} D_{ij}=GdG^{*}\cap
G^{*}dG=C(d)\cap c(d)  \ \  . \end{equation}
The tangent space to $D_{ij}$ may be represented
as an intersection of  tangent spaces to left
and right cells $C(d)$ and $c(d)$:
\begin{equation} T_{d}^{'}=T_{d}(C(d))\cap T_{d}(c(d))
\ \  .  \end{equation}
For the latter the following formulae are true:
\begin{equation}  T_{d}(C(d))=\g+Ad(d)\g^{*} \ \
,  \end{equation}
\begin{equation}  T_{d}(c(d))=\g^{*}+Ad(d)\g  \ \
. \end{equation}
The space $T_{d}(C(d))$ coincides with $V(d)^{\perp}$.
Indeed, $T_{d}(C(d))^{\perp}$ lies in $\g$
because $T_{d}(C(d))^{\perp}\!\subset\! \g^{\perp}\!
=\!\g$.
On the other hand
\begin{equation}
<T_{d}(C(d))^{\perp},Ad(d)\g^{*}>_{\d}=0 \ \
. \end{equation}
Formula (104) implies
that $Ad(d^{-1})T_{d}(C(d))^{\perp}\!\subset\!
\g^{*\perp}\! =\!\g^{*}$. So $T_{d}(C(d))^{\perp}$
is the subspace in $\g$ which is mapped by $Ad(d^{-1})$
into $\g^{*}$. It is the subspace $V(d)$
that satisfies these conditions. So we have
\begin{equation}  T_{d}(C(d))^{\perp}=V(d)\ \
,\ \ T_{d}(C(d))=V(d)^{\perp}  \ \  .  \end{equation}

Similarly,
\begin{equation}  T_{d}(c(d))=V^{*}(d)^{\perp} \ \
.  \end{equation}

Comparing (99), (101), (105), (106), we conclude
that the tangent space $T_{d}^{'}$
to the cell $D_{ij}$ coincides with
the tangent space $T_{d}^{S}$ to the symplectic leaf.
Thus the symplectic leaf coincides
with a connected component of the cell $D_{ij}$.

We have proved theorem 2. The next question concerns
the symplectic structure on the leaves $D_{ij}$.

\section{Symplectic structure of the Heisenberg
\mbox{double.}}

Each symplectic leaf $D_{ij}$ introduced
in the last section carries a nondegenerate
Poisson structure
and hence the corresponding symplectic form $\O_{ij}$
can be defined. To write down the answer
we need several new objects. Let us denote
by $L_{ij}$ the subset in $G\times G^{*}$ defined
as follows:
\begin{equation}  L_{ij}=\{ (g,g^{*})\in
G\times G^{*}\ \ ,\ \ gd_{i}g^{*}\in D_{ij}\}
\ \  . \end{equation}
In the same way we construct the subset $M_{ij}$
in $G^{*}\times G$:
\begin{equation}  M_{ij}=\{ (h^{*},h)\in
G^{*}\times G\ \ ,\ \ h^{*}d_{j}^{-1}h\in D_{ij}\}
\ \  . \end{equation}
Finally let $N_{ij}$ be the subset
in $L_{ij}\times M_{ij}$:
\begin{equation}
N_{ij}=\{ [(g,g^{*}),(h^{*},h)]\in L_{ij}\times
M_{ij}\ \ ,\ \ gd_{i}g^{*}=h^{*}d_{j}^{-1}h\} \ \   .
\end{equation}
We can define the projection
\begin{equation}
p_{ij}\ \ :\ \ N_{ij}\longrightarrow D_{ij}
\end{equation}
\begin{equation}
p_{ij}\ \ :\ \ [(g,g^{*}),(h^{*},h)]\longrightarrow
d=gd_{i}g^{*}=h^{*}d_{j}^{-1}h   \end{equation}
and consider the form $p_{ij}^{*}\O_{ij}$
on $N_{ij}$ instead of the original form $\O_{ij}$
on $D_{ij}$. It is parallel to the construction
of the Kirillov form
on the coadjoint orbit (see section 1).
Parametrizations (107), (108) provide us
with the coordinates $(g,g^{*})$ and $(h^{*},h)$
on $N_{ij}$. We can use them to write down the answer:

\begin{fact}
The symplectic form $p_{ij}^{*}\O_{ij}$ on $N_{ij}$
can be represented as follows:
\begin{equation}   p_{ij}^{*}\O_{ij}=\frac{1}{2}
(<\o_{h^{*}}\!\stackrel{\wedge}{,}\o_{g}\! >
+<\m_{g^{*}}\!\stackrel{\wedge}{,}\m_{h}\! >) \ \
.   \end{equation}
\end{fact}

In the formula (112) $\o_{g},\o_{h^{*}}
,\m_{h},\m_{g^{*}}$ are restrictions
of the corresponding one-forms from
$(G\times G^{*})\times (G^{*}\times G)$
to $N_{ij}$. The pairing $<\, ,>$ is applied
to values of  Maurer-Cartan forms,
which can be treated as elements of $\g$
and $\g^{*}$ embedded to $\d =\g +\g^{*}$.
So we can use $<\, ,>_{\d}$ as well as $<\, ,>$.

{\em Proof of theorem 3.}

The strategy of the proof is quite straightforward.
We consider Poisson bracket (79)
on the symplectic leaf $D_{ij}$.
If we use dual bases $\{ e_{a}\} $
and $\{ e^{a}\} $ $(a=1,\ldots ,n\! =\! dimD)$
of right-invariant vector fields
and one-forms on $D$, the formula (79) acquires
the following form:
\begin{eqnarray}  \{ f,h\} (d)=-<\nabla \!
\! _{L}f\otimes \nabla \! \! _{L}h,r+Ad(d)\otimes
Ad(d)\, r^{*}>=  \nonumber \\ =-\sum_{a,b=1}^{n}<\nabla
\! \! _{L}f,e_{a}><\nabla \! \! _{L}h,e_{b}><e^{a},\P
Je^{b}> \ \  .
\end{eqnarray}
The last multiplier in (113) is Poisson matrix
corresponding to the bracket (79):
\begin{equation}  \P^{ab}=<e^{a},\P Je^{b}> \ \
. \end{equation}
Here $\P$ is the same as in (86).
The matrix $\P^{ab}$ may be degenerate.
Let us choose vectors $\{ e_{a}\ \ ,\
\ a\in s_{ij}=\{1,\ldots ,n_{ij}\! =\!
dimD_{ij}\} \} $ so that they form a basis
in the space $T_{d}$ tangent to $D_{ij}$. $\P^{ab}$
is not zero only if both $a$ and $b$ belong to $s_{ij}$.
The symplectic form $\O_{ij}$ on the cell $D_{ij}$
can be represented as follows (see section 1):
\begin{equation}
\O_{ij}=\sum_{a,b=1}^{n_{ij}}\O_{ab}e^{a}\!
\!\otimes\! e^{b} \ \ ,  \end{equation}
where the matrix $\O$ satisfies the following condition:
\begin{equation}
\sum_{c=1}^{n{ij}}\O_{ac}\P^{cb}=\delta_{a}^{b}
\ \  . \end{equation}
So what we need is inverse matrix $\P^{-1}$
for $\P^{ab}$.
To make the symbol $\P^{-1}$ meaningful
we introduce two operators $\P_{1}$ and $\P_{2}$:
\begin{equation}  \P_{1}=(P+Ad(d)P^{*}) \ \
, \end{equation}
\begin{equation}  \P_{2}=(P^{*}-Ad(d)P)  \ \
. \end{equation}
$\P$ may be decomposed in two ways,
using $\P_{1}$ and $\P_{2}$:
\begin{equation}
\P=\P_{1}\P_{2}^{*}=-\P_{2}\P_{1}^{*}
\ \ . \end{equation}
Some useful properties
of the operators $\P_{1}$ and $\P_{2}$
are collected in the following lemma.

\begin{eqnarray} {\bf Lemma \ \  1}\ \ \ \ \ \ \
\ \ \ \ \ \ \ \ \ \ \ \ \ \ \ \
\  Im\P_{1}=V(d)^{\perp}\ \ ,\
\ Im\P_{2}=V^{*}(d)^{\perp}\ \ ,\ \ \nonumber  \\
\ \ \
P(Ker\P_{1})=V(d)\ \ ,\ \ P^{*}(Ker\P_{2})=V^{*}(d)
\ \ .   \end{eqnarray}
{\em Proof.}  First let us consider the formula
\begin{equation}  Im\P_{1}=(Ker\P_{1}^{*})^{\perp} \ \
. \end{equation}
The operator $\P_{1}^{*}$ looks like follows:
\begin{equation}  \P_{1}^{*}=P^{*}+PAd(d^{-1}) \ \
. \end{equation}
The equation for $Ker\P_{1}^{*}$
\begin{equation}  (P^{*}+PAd(d^{-1}))(x+\e)=0
\end{equation}
leads immediately to the following restrictions
for $x$ and $\e$:
\begin{equation} x=0\ \ ,\ \ Ad(d^{-1})\e \in \g^{*}
\ \  .  \end{equation}
Comparing (124) with definition (95),
we see that $Ker\P_{1}^{*}=V(d)$
and hence $Im\P_{1}=V(d)^{\perp}$.

If a vector $x+\e$ belongs
to the kernel of the operator $\P_{1}$,
it satisfies the following equation:
\begin{equation} (P+Ad(d)P^{*})(x+\e )=0 \ \
.  \end{equation}
It can be rewritten as a set of conditions
for the components $x,\e$:
\begin{equation} Ad(d^{-1})\e \in \g^{*}\ \
,\ \ x=-Ad(d^{-1})\e \ \  .  \end{equation}
$\e$ again appears to be an element of $V(d)$.
This fact may be represented
as the equation $P(Ker\P_{1})=V(d)$.

We omit the proofs of the formulae (120)
concerning the operator $\P_{2}$
because they are parallel to the proofs given above.

The following step is to define inverse operators:
\begin{equation}  \P_{1}^{-1}\ \ :
\ \ Im\P_{1}\longrightarrow \d /Ker\P_{1}\ \
,  \end{equation}   \begin{equation}
\P_{2}^{-1}\ \ :\ \ Im\P_{2}\longrightarrow
\d /Ker\P_{2} \ \  . \end{equation}
The solution of the equation
\begin{equation}  \P_{1,2}^{-1}a=b   \end{equation}
exists if and only if $a\in Im\P_{1,2}$ and $b$
is defined up to an arbitrary vector from
$Ker\P_{1,2}$.

Now we are ready to write down the answer
for $\O_{ab}$:
\begin{equation}  \O_{ab}=<e_{a},\O e_{b}>_{\d}
\ \  ,\ \  \O=P\P_{1}^{-1}-P^{*}\P_{2}^{-1}  \ \
. \end{equation}
First of all let us check
that matrix elements $\O_{ab}$ are well-defined.
Vectors $e_{b}$  form the basis
in the space $T_{d}=(V(d)\oplus V^{*}(d))^{\perp}$.
Both $\P_{1}^{-1}$ and $\P_{2}^{-1}$
are defined on $T_{d}$
because $T_{d}\subset V(d)^{\perp}=Im\P_{1}$
and also $T_{d}\subset V^{*}(d)^{\perp}=Im\P_{2}$.
So the vector $\O e_{b}$ exists but it is not unique.
It is defined up to an arbitrary vector
\begin{equation}
\delta \in P(Ker\P_{1})+P^{*}(Ker\P_{2})=V(d)+V^{*}(d)
\ \  .  \end{equation}
Fortunately
the vector $e_{a}\!\in\! T_{d}$
and $<e_{a},\delta >=0$ for any $\delta$
of the form (131). We conclude
that the ambiguity in the definition
of the operator $\O$ does not
lead to an ambiguity for matrix elements $\O_{ab}$.

Now we must check condition (116):
\begin{eqnarray}
\delta_{a}^{b}=\sum_{c=1}^{n_{ij}}\O_{ac}\P^{cb}=\
\ \ \ \ \  \ \ \ \ \ \ \ \nonumber \\
=\sum_{c=1}^{n_{ij}}<e_{a},\O
e_{c}><e^{c},\P J(e^{b})>= \\
 =<e_{a},\O \P J(e^{b})>_{\d}   \ \  .\
\ \ \ \ \ \ \ \ \ \nonumber   \end{eqnarray}
The product $\O \P $ can be easily calculated
using (119),(130):
\begin{eqnarray}
\O \P =P\P_{1}^{-1}\P_{1}\P_{2}^{*}
+P^{*}\P_{2}^{-1}\P_{2}\P_{1}^{*}=\ \ \ \ \ \ \
\nonumber \\
=P(P-P^{*}Ad(d^{-1}))+P^{*}(P^{*}+PAd(d^{-1}))= \\
=P+P^{*}=I \ \  . \ \ \ \ \ \ \ \ \ \ \ \ \ \ \
\ \ \ \ \ \ \ \nonumber \end{eqnarray}
We must remember that the vector $\O \P J(e^{b})$
is defined up to an arbitrary vector from
$V(d)\oplus V^{*}(d)$
because we used in (133) the ``identities''
\begin{equation}
\P_{1}^{-1}\P_{1}\approx\P_{2}^{-1}\P_{2}\approx
id \ \  .  \end{equation}
The ambiguity in (134) does not influence
 the answer:
\begin{equation} <e_{a},\O
\P J(e^{b})>_{\d}=<e^{b},e_{a}>=\delta_{a}^{b}
\end{equation}
as it is required by (116).

We can rewrite formula (130)
in more invariant way:
\begin{equation}
\O_{ij}=<\o_{d}^{ij}\stackrel{\otimes}{,}\O
\o^{ij}_{d}>_{\d} \ \  , \end{equation}
where $\o^{ij}_{d}$ is the restriction
of the Maurer-Cartan form to the cell $D_{ij}$.
Expression (130) for the operator $\O$
still includes inverse  operators $\P^{-1}_{1,2}$
implying that some equations must be solved.
To this end we consider the pull-back
of the form $\O_{ij}$:
\begin{equation}
p_{ij}^{*}\O_{ij}=<p_{ij}^{*}\o^{ij}_{d}
\stackrel{\otimes}{,}\O p_{ij}^{*}\o^{ij}_{d}>_{\d}
\ \  . \end{equation}
There are coordinates $(g,g^{*})$ and $(h^{*},h)$
on $N_{ij}$.
The Maurer-Cartan form $p_{ij}^{*}\o^{ij}_{d}$
can be rewritten in two ways:
\begin{equation} p_{ij}^{*}\o^{ij}_{d}=\o_{g}
+Ad(d)\m_{g^{*}} \ \  ,  \end{equation}
\begin{equation} p_{ij}^{*}\o^{ij}_{d}=\o_{h^{*}}
+Ad(d)\m_{h}  \ \  .  \end{equation}
Representations (138), (139) allow us
to calculate $\P_{1,2}^{-1}p_{ij}^{*}\o^{ij}_{d}$
explicitly:
\begin{equation}
\P_{1}^{-1}p_{ij}^{*}\o^{ij}_{d}=\o_{g}
+\m_{g^{*}} \ \  ,  \end{equation}
\begin{equation}
\P_{2}^{-1}p_{ij}^{*}\o^{ij}_{d}=\o_{h^{*}}
-\m_{h}  \ \ . \end{equation}
Let us mention again that solutions (140), (141)
are not unique.
We can take any possible value of $\O \o_{d}^{ij}$.
The answer for the form $\O_{ij}$ is independent
of this choice.

Putting together (130), (137), (140) and (141),
we obtain the following formula
for the symplectic form:
\begin{eqnarray}
p_{ij}^{*}\O_{ij}=<(\o_{g}
+Ad(d)\m_{g^{*}})\stackrel{\otimes}{,}\o_{g}>_{\d}
-<(\o_{h^{*}}
+Ad(d)\m_{h})\stackrel{\otimes}{,}\o_{h^{*}}>_{\d}=
\nonumber  \\
=<Ad(d)\m_{g^{*}}\stackrel{\otimes}{,}\o_{g}>_{\d}
-<Ad(d)\m_{h}\stackrel{\otimes}{,}\o_{h^{*}}>_{\d}
\ \  .  \ \ \ \ \ \ \ \end{eqnarray}
Actually, the form (142) is antisymmetric.
To make it evident, let us consider the identity
\begin{eqnarray}  <p_{ij}^{*}\o_{d}^{ij}
\stackrel{\otimes}{,}p_{ij}^{*}\o_{d}^{ij}>_{\d}=
\ \ \ \ \ \ \ \ \ \ \ \ \ \ \ \ \ \ \nonumber \\
=<Ad(d)\m_{g^{*}}\stackrel{\otimes}{,}\o_{g}>_{\d}
+<\o_{g}\stackrel{\otimes}{,}Ad(d)\m_{g^{*}}>_{\d}
= \ \    \\
=<Ad(d)\m_{h}\stackrel{\otimes}{,}\o_{h^{*}}>_{\d}
+<\o_{h^{*}}\stackrel{\otimes}{,}Ad(d)\m_{h}>_{\d}
\ \ . \nonumber  \end{eqnarray}
Or, equivalently,
\begin{eqnarray}  <Ad(d)\m_{g^{*}}
\stackrel{\otimes}{,}\o_{g}>_{\d}-<Ad(d)\m_{h}
\stackrel{\otimes}{,}\o_{h^{*}}>_{\d}=
\ \ \ \  \nonumber  \\
=-<\o_{g}\stackrel{\otimes}{,}Ad(d)\m_{g^{*}}>_{\d}
+<\o_{h^{*}}\stackrel{\otimes}{,}Ad(d)\m_{h}>_{\d}
\ \ . \end{eqnarray}
Applying (144) to make (142) manifestly antisymmetric,
one gets:
\begin{equation}
p_{ij}^{*}\O_{ij}=\frac{1}{2}(<Ad(d)\m_{g^{*}}
\stackrel{\wedge}{,}\o_{g}>_{\d}+
<\o_{h^{*}}\stackrel{\wedge}{,}Ad(d)\m_{h}>_{\d})
\ \  .  \end{equation}

Using representation (111) of $d$ in terms
 of $(g,g^{*})$ and $(h^{*},h)$,
it is easy to check that formula (145)
coincides with
\begin{equation}
p_{ij}^{*}\O_{ij}=-\frac{1}{2}(<\m_{g}
\stackrel{\wedge}{,}Ad(d_{i})\o_{g^{*}}>_{\d}+
<\o_{h}\stackrel{\wedge}{,}Ad(d_{j})\m_{h^{*}}>_{\d})
\ \ .  \end{equation}
To obtain formula (112) one can use (138),(139):
\begin{equation}  p_{ij}^{*}\o^{ij}_{d}=\o_{g}
+Ad(d)\m_{g^{*}}=\o_{h^{*}}+Ad(d)\m_{h} \ \
. \end{equation}
Or, equivalently,
\begin{equation} \o_{g}-Ad(d)\m_{h}=\o_{h^{*}}
-Ad(d)\m_{g^{*}} \ \  .  \end{equation}
Due to antisymmetry we have
\begin{equation} <(\o_{g}-Ad(d)\m_{h})
\stackrel{\wedge}{,}(\o_{h^{*}}
-Ad(d)\m_{g^{*}})>_{\d}=0 \ \ .  \end{equation}
Therefore,
\begin{eqnarray}  \frac{1}{2}(<\o_{h^{*}}
\stackrel{\wedge}{,}\o_{g}>_{\d}+<\m_{g^{*}}
\stackrel{\wedge}{,}\m_{h}>_{\d})=
\ \ \ \ \ \ \ \ \ \ \ \ \ \nonumber  \\
=\frac{1}{2}(<Ad(d)\m_{g^{*}}
\stackrel{\wedge}{,}\o_{g}>_{\d}+<\o_{h^{*}}
\stackrel{\wedge}{,}Ad(d)\m_{h}>_{\d})
=p_{ij}^{*}\O_{ij} \ \ ,\end{eqnarray}
which coincides with (112).

Now we have to check that the r.h.s.
of formula (112) does represent the pull-back
of some two-form on $D_{ij}$.
The problem is in the ambiguity of formula (70).
Coordinates $g$ and $g^{*}$ are defined
only up to the following change of variables:
\begin{equation}  g^{'}=gs\ \ ,\ \ g^{*'}=tg^{*}
\ \ , \end{equation}
where
\begin{equation}  sd_{i}t=d_{i}  \ \ .
\end{equation}
Here $s$ is an element of $S(d_{i})$
and $t$ belongs to $T(d_{i})$.
The parameter $s$ determines $t$
by means of formula (152).
Similar ambiguity exists in the definition of $h$
and $h^{*}$.
We can construct an infinitesimal analogue
of formula (151). The vector field $v_{\e}$
on $N_{ij}$
\begin{equation}  v_{\e}=(Ad(g)\e, -Ad(d_{i}^{-1})\e)
\ \   \end{equation}
does not correspond to any nonzero vector field
on $D_{ij}$.
Here we use coordinates $(g,g^{*})$ on $N_{ij}$
and left identification of vector fields
on $G\times G^{*}$ and $\g +\g^{*}$.
So the first term is an element of $\g$
and the second one belongs to $\g^{*}$.
Therefore $Ad(g)\e$ belongs to $V(d_{i})$
(see section 2).

Actually we must check two nontrivial statements:
\begin{enumerate}
\item Form $p_{ij}^{*}\O_{ij}$ is invariant
with respect to change of variables (151).
It follows from the definition
of the Maurer-Cartan forms $\o$ and $\m$.
\item Tangent vectors (153)
belong to the kernel of $p_{ij}^{*}\O_{ij}$.
\end{enumerate}
It is convenient to use expression (146)
for $p_{ij}^{*}\O_{ij}$:
\begin{eqnarray}
p_{ij}^{*}\O_{ij}=-\frac{1}{2}(<\m_{g}
\stackrel{\wedge}{,}Ad(d_{i})\o_{g^{*}}>_{\d}+
<\o_{h}
\stackrel{\wedge}{,}Ad(d_{j})\m_{h^{*}}>_{\d})=
\nonumber \\
=-\frac{1}{2}(\omega_{1}+\omega_{2}) \ \
,\ \ \ \ \ \ \ \ \ \ \ \ \ \ \ \ \ \ \ \ \ \
\end{eqnarray}
where
\begin{equation} \omega_{1}=<\m_{g}
\stackrel{\wedge}{,}Ad(d_{i})\o_{g^{*}}>_{\d}
\ \  ,  \end{equation}
\begin{equation} \omega_{2}=<\o_{h}
\stackrel{\wedge}{,}Ad(d_{j})\m_{h^{*}}>_{\d}
\ \ .  \end{equation}
We have to consider $\omega_{1}(\ .\ ,v_{\e})$
and $\omega_{2}(\ .\ ,v_{\e})$.
\begin{eqnarray}  \omega_{1}(\ .\ ,v_{\e})
=<\m_{g},Ad(d_{i})\o_{g^{*}}(v_{\e})>_{\d}-
<\m_{g}(v_{\e}),Ad(d_{i})\o_{g^{*}}>_{\d}=
\ \ \ \nonumber  \\
=<\m_{g},Ad(d_{i})Ad(d_{i}^{-1})\e >_{\d}
+<Ad(g^{-1})Ad(g)\e ,Ad(d_{i})\o_{g^{*}}>_{\d}=
\\
=<\m_{g},\e >_{\d}+<\o_{g^{*}},Ad(d_{i}^{-1})\e>_{\d}
\ \  .\ \ \ \ \ \ \ \ \ \ \ \ \ \ \ \ \ \ \ \ \nonumber
\end{eqnarray}
Here we use properties (16), (17)
of the Maurer-Cartan forms.
It is easy to see that both terms
in the last expression (157) are equal to zero.
First of them
\begin{equation}  <\m_{g},\e>_{\d}=0
\end{equation}
because both $\e$ and a value of $\m_{g}$
belong to $\g$. All the same with the second term:
\begin{equation}  <\o_{g^{*}},Ad(d_{i}^{-1})\e>_{\d}=0
\end{equation}
because for $Ad(g)\e \in V(d_{i})$
the combination $Ad(d^{-1}_{i})\e $ belongs to $\g^{*}$.
We remind that both $\g$ and $\g^{*}$
are isotropic subspaces in $\d$.

We omit the proof for the second term $\omega_{2}$
in (154) because it is quite  parallel to the one
described above. We conclude
that form (112) indeed corresponds to some two-form
on the symplectic leaf $D_{ij}$.

It is known from general Poisson theory that
\begin{equation}   d\O =0 \ \ , \end{equation}
but it is interesting to check that form (112)
is closed by direct calculations.
Rewriting equation (148) we get:
\begin{equation}  \o_{g}-\o_{h^{*}}
=Ad(d)\m_{h}-Ad(d)\m_{g^{*}} \ \ . \end{equation}
Taking the cube of the last equation we get:
\begin{eqnarray}  <\o_{g}
\stackrel{\wedge}{,}\o_{g}\wedge\o_{g}>_{\d}
-<\o_{h^{*}}\stackrel{\wedge}{,}\o_{h^{*}}
\wedge\o_{h^{*}}>_{\d}+ \ \ \ \ \  \nonumber  \\
+3<\o_{g}\stackrel{\wedge}{,}\o_{h^{*}}
\wedge\o_{h^{*}}>_{\d}-3<\o_{g}\wedge\o_{g}
\stackrel{\wedge}{,}\o_{h^{*}}>_{\d}=
\ \ \ \nonumber  \\
=<\m_{h}\stackrel{\wedge}{,}\m_{h}\wedge\m_{h}>_{\d}
-<\m_{g^{*}}\stackrel{\wedge}{,}\m_{g^{*}}
\wedge\m_{g^{*}}>_{\d} +  \ \ \\
+3<\m_{h}\stackrel{\wedge}{,}\m_{g^{*}}
\wedge\m_{g^{*}}>_{\d}-3<\m_{h}\wedge\m_{h}
\stackrel{\wedge}{,}\m_{g^{*}}>_{\d}
\ \ .\nonumber \end{eqnarray}
As $\o_{g}\wedge\o_{g}=\frac{1}{2}[\o_{g}
\stackrel{\wedge}{,}\o_{g}]$ and
$\m_{h}\wedge\m_{h}=\frac{1}{2}[\m_{h}
\stackrel{\wedge}{,}\m_{h}]$
take values in $\g$, $\o_{h^{*}}\wedge\o_{h^{*}}
=\frac{1}{2}[\o_{h^{*}}\stackrel{\wedge}{,}\o_{h^{*}}]$
and $\m_{g^{*}}\wedge\m_{g^{*}}=\frac{1}{2}[\m_{g^{*}}
\stackrel{\wedge}{,}\m_{g^{*}}]$ take values in $\g^{*}$
we may use the pairing $<\, ,>_{\d}$ for them.
Moreover, as both $\g$ and $\g^{*}$
are isotropic subspaces in $\d$, we rewrite (162)
as follows:
\begin{eqnarray} <\o_{g}\stackrel{\wedge}{,}\o_{h^{*}}
\wedge\o_{h^{*}}>_{\d}-<\o_{g}\wedge\o_{g}
\stackrel{\wedge}{,}\o_{h^{*}}>_{\d}-
\ \ \ \ \  \nonumber  \\
-<\m_{h}\stackrel{\wedge}{,}\m_{g^{*}}
\wedge\m_{g^{*}}>_{\d}+<\m_{h}\wedge\m_{h}
\stackrel{\wedge}{,}\m_{g^{*}}>_{\d}=0
\ \  . \end{eqnarray}
We remind that $d\o_{g}=\o_{g}\wedge\o_{g}$
and $d\m_{g}=-\m_{g}\wedge\m_{g}$. Thus,
\begin{eqnarray} dp_{ij}^{*}\O_{ij}
=-<d\o_{g}\stackrel{\wedge}{,}\o_{h^{*}}>_{\d}
+<\o_{g}\stackrel{\wedge}{,}d\o_{h^{*}}>_{\d}-
\nonumber  \\
-<d\m_{h}\stackrel{\wedge}{,}\m_{g^{*}}>_{\d}
+<\m_{h}\stackrel{\wedge}{,}d\m_{g^{*}}>_{\d}=0
\ \ .  \end{eqnarray}

Now it is interesting to consider the classical limit
of our theory
to recover the standard answer for $T^{*}\!G$.
There is no deformation parameter in  bracket (79)
but it may be introduced by hand:
\begin{equation}  \{ f,h\}_{\gamma}=\gamma \{ f,h\}
\ \  . \end{equation}
For the new bracket (165) we have the symplectic form:
\begin{equation}
\O_{ij}^{\gamma}=\frac{1}{\gamma}\O_{ij}
\ \ .  \end{equation}
The classical limit $\gamma \rightarrow 0$ makes sense
only for the main cell corresponding to $d_{i}=d_{j}=I$.
The idea is to parametrize a vicinity of the unit element
in the group $G^{*}$ by means of the exponential map:
\begin{equation}  g^{*}=\exp(\gamma m) \ \  , \end{equation}
\begin{equation}  h^{*}=\exp(\gamma l)  \ \ , \end{equation}
where $m$ and $l$ belong to $\g^{*}$.
Coordinates $m$ and $l$ are adjusted in such a way
that they have finite values after the limit procedure.
When $\gamma $ tends to zero, the formula
\begin{equation}  d=gg^{*}=h^{*}h  \end{equation}
leads to the following relations:
\begin{equation} g=h \ \ ,\ \ l=Ad^{*}(g)m
\ \  .  \end{equation}
Expanding the form $\O^{\gamma}$
into the series in $\gamma $ we keep only the constant term
(singularity $\gamma^{-1}$ disappears from the answer
because the corresponding two-form
is identically equal to zero). The answer is the following:
\begin{equation}  \O^{\gamma}
=\frac{1}{2}(<dm\stackrel{\wedge}{,}\m_{g}>
+<dl\stackrel{\wedge}{,}\o_{g}>) \end{equation}
and it recovers classical answer (49) (see section 1).
Deriving formula (171), we use the expansions
for the Maurer-Cartan forms on $G^{*}$:
\begin{equation} \o_{g^{*}}=\gamma dm+O(\gamma^{2})
\ \ ,      \end{equation}
\begin{equation} \m_{h^{*}}=\gamma dl+O(\gamma^{2})
\ \  .     \end{equation}

We have considered general properties
of the symplectic structure
on the Heisenberg double $D_{+}$ and now we turn
to the theory of orbits for Lie-Poisson groups.

\section{Theory of orbits.}

In this section we describe reductions
of the Heisenberg double $D_{+}$
which lead to Lie-Poisson analogues of coadjoint orbits.
We consider quotient spaces of the double D
over its subgroups $G$ and $G^{*}\ $:$\ \ F_{R}\!
=\! D\! /\! G\ ,\   \ F_{R}^{*}\! =\! D\! /\! G^{*}\
,\ \ F_{L}\! =\! G\!\setminus\! D\ ,\\ F_{L}^{*}\!
=\! G^{*}\!\setminus \! D$. They inherit Poisson
bracket from the double $D_{+}$.
Indeed, let us pick up $F_{R}$ as an example.
Functions on $F_{R}$ may be regarded
as functions on $D$ invariant with respect
to right action of $G\,$:
\begin{equation} f(dg)=f(d) \ \ .
\end{equation}
The right derivative $\nabla \! \! _{R}f$
is orthogonal to $\g\,$ for functions on $F_{R}$:
\begin{equation} <\nabla \! \! _{R}f,\g>=0 \ \ .
\end{equation}
For a pair of invariant functions $f$
and $h$ the second term in the formula (79) vanishes
because $r^{*}\!\in\! \g^{*}\!\otimes\! \g$.
The first term is an invariant function because
the left derivative $\nabla\!\! _{L}$
preserves the condition (174).
So we conclude that the Poisson bracket
\begin{equation} \{ f,h\} =-<\nabla \!
\! _{L}f\otimes \nabla \! \! _{L}h,r>
\end{equation}
is well-defined on invariant functions
and hence it can be treated as a Poisson bracket
on $F_{R}$.
The purpose of this section is to study
the stratification of the space $F_{R}$
into symplectic leaves and describe
the corresponding symplectic forms on them.
One can consider $F_{L},\ F_{R}^{*},\ F_{L}^{*}$
in the same way.

Using stratification (77) of the double $D$
we can obtain the stratification of the space $F_{R}$:
\begin{equation}
F_{R}=\bigcup_{j}G^{*}\! /T_{-j}=\bigcup_{j}G^{*}_{j}
\ \  .  \end{equation}
Each stratification cell $G_{j}^{*}$
is just an orbit of the natural action of $G^{*}$
on the quotient space $F_{R}=D\! /\! G$
by the left multiplication.
We denote the orbit of the class of unity in $D$
by $G_{0}^{*}$. It is a quotient of $G^{*}$
over discrete subgroup $\cal E$$=\! G^{*}\!\cap\! G,\
G_{0}^{*}\!=\!G\! /\!$$\cal E$.

We have factorized the double $D$
over the right action of the group $G$.
However, the same group acts on the quotient space
by the left multiplications:
\begin{equation}  g\ \ :\ \ dG\longrightarrow gdG
\ \  .  \end{equation}
Here the class $dG$ is mapped into the class $gdG$.
In the vicinity of the unit element on the maximum cell
\mbox{$GG^{*}\cap G^{*}G$} the action (178)
looks like follows:
\begin{equation}
gg^{*}=g^{*'}\! (g,g^{*})g^{'}(g,g^{*})
\ \ .  \end{equation}
The element $g^{*'}\!(g,g^{*})$
is a result of the left action of the element $g$
on the point $g^{*}\!\!\in\! G^{*}\!\!\subset\! F_{R}$.
 In the classical limit, when $g^{*}$ and $g^{*'}$
are very close to the identity, formula (179)
transforms into the coadjoint action of $G$ on $\g^{*}$:
\begin{equation}  g^{*}=I+\gamma l+\ldots
\ \  ,  \end{equation}
\begin{equation}  g^{*'}=I+\gamma l^{'}
+\ldots \ \  ,  \end{equation}
\begin{equation}  l^{'}=Ad^{*}(g)l
\ \  .  \end{equation}
For historical reasons transformations (179)
are called dressing transformations.
We denote them $AD^{*}$ to remind their relation
to the coadjoint action:
\begin{equation}  g^{*'}\! (g,g^{*})=AD^{*}(g)g^{*}
\ \  .  \end{equation}

As we have mentioned, the transformation $AD^{*}$
is defined on the space $F_{R}$ globally.
For some values of $g$ and $g^{*}$ in (183)
the element $g^{*'}$ does not exist
and the result of the action of $g$ on $g^{*}$
belongs to some other cell $G_{j}^{*}$
of stratification (177). So we have a correct definition
of the $AD^{*}$-orbit in the Lie-Poisson case.
The question is whether they coincide
with symplectic leaves or not.
In general the answer is negative.
Characterizing the situation we shall systematically omit
the proofs concerning standard Poisson theory \cite{2,3}.

A powerful tool for studying symplectic leaves
is a dual pair. By definition
a pair of Poisson mappings of symplectic manifold $S$
to different Poisson manifolds $P_{L}$ and $P_{R}$:
\begin{eqnarray}  S \ \ \ \ \ \ \ \ \ \ \ \ \ \
\nonumber  \\
\swarrow  \ \ \ \  \searrow \ \ \ \ \ \ \ \ \  \\
P_{L} \ \ \ \ \ \ \   P_{R} \ \ \ \ \ \ \
\nonumber  \end{eqnarray}
is called a dual pair, if Poisson bracket
of any function on $S$ lifted from $P_{R}$
vanishes when the second function is lifted from $P_{L}$
and in this case only.
Symplectic leaves in $P_{R}$
can be obtained in the following way.
Take a point in $P_{L}$, consider its preimage in $S$
and project it into $P_{R}$. Connected components
of the image of this projection
are symplectic leaves in $P_{R}$.

As an example let us consider
the following pair of Poisson mappings:
\begin{eqnarray}  D_{+} \ \ \ \ \ \ \ \ \ \ \ \
\nonumber  \\
\swarrow  \ \ \ \  \searrow \ \ \ \ \ \ \ \ \  \\
F_{L} \ \ \ \ \ \ \ \   F_{R} \ \ . \ \ \ \
\nonumber  \end{eqnarray}
This pair is not a dual pair because $D_{+}$
is not a symplectic manifold. However,
the pair (185) is related to a family of dual pairs:
\begin{eqnarray}  D_{ij} \ \ \ \ \ \ \ \ \ \ \ \
\nonumber  \\
\swarrow  \ \ \ \  \searrow \ \ \ \ \ \ \ \ \  \\
F_{L} \ \ \ \ \ \ \ \  F_{R} \ \ . \ \ \ \ \nonumber
\end{eqnarray}
Here we use symplectic leaves $D_{ij}$
instead of $D_{+}$. One can prove
that pair of mappings (186) is a dual pair
by direct calculation with bracket (79).
Choosing dual pairs
with different indices $\scriptstyle ij$,
we cover all space $F_{R}$
and find all the symplectic leaves in this space.

Let us apply the general prescription
to the dual pair (186).
We pick up a class $Gx\in
Im\! _{\scriptscriptstyle L}D_{ij}\subset T_{i}\!
\!\setminus\!\! G^{*}\subset F_{L}$.
Its preimage in $D_{ij}$ is an intersection $ K_{ij}(x)
=Gx\cap D_{ij}$. Projecting $K_{ij}(x)$ into $F_{R}$,
we get a symplectic leaf:
\begin{equation}  AD^{*}(G)xG\cap
Im\! _{\scriptscriptstyle R}D_{ij}
\ \  .  \end{equation}
Let us remark that $Im\! _{\scriptscriptstyle R}D_{ij}$
is an intersection $G_{j}^{*}\cap (\cup_{g^{*}\in G^{*}}
AD^{*}(G)d_{i}g^{*}G)$. It implies that
we may use $G_{j}^{*}$
instead of $Im\! _{\scriptscriptstyle R}D_{ij}$
in the formula (187).
So all the symplectic leaves in $F_{R}$
are  intersections of orbits
of  dressing transformations $AD^{*}$
and orbits $G^{*}_{j}$ of the action of $G^{*}$
in $F_{R}$. To get all the leaves
we have to use all the cells $D_{ij}$ in $D$.
The orbits of $AD^{*}$-action in $F_{R}$
appear to have a complicated structure.
Each orbit $O_{p_{0}}=AD^{*}(G)\, p_{0}\ $
$(\ p_{0}\in F_{R})$
may be represented as a sum of its cells:
\begin{equation}  O_{p_{0}}=\bigcup_{j}(AD^{*}(G)\,
p_{0}\cap G^{*}_{j})=\bigcup_{j}O^{j}_{p_{0}}
\ \ . \end{equation}
Each cell of stratification (188)
is a symplectic leaf in $F_{R}$.

Now we turn to the description of symplectic forms
on the leaves (188). As usually, it is convenient
to use coordinates on the orbit and on the group $G$
at the same time. Formula
\begin{equation}  gh^{*}_{0}d_{j}^{-1}G=h^{*}d_{j}^{-1}G
\end{equation}
for the action of $AD^{*}$
on the point $h_{0}^{*}T_{-j}\in G^{*}_{j}$ provides us
with the projection from the subset
\begin{equation}  G_{j}(h_{0}^{*})=\{ g\in G\ \
,\ \ gh_{0}^{*}d_{j}^{-1}\in G^{*}d_{j}^{-1}G\}
\end{equation}
to the cell $O^{j}_{h_{0}^{*}}$ of the orbit:
\begin{equation}
p_{j}\ \ :\ \ G_{j}(h_{0}^{*})\longrightarrow
O^{j}_{h_{0}^{*}} \ \  , \end{equation}
\begin{equation}
p_{j}\ \ :\ \ g\longrightarrow h^{*}T_{-j}
\ \  , \end{equation}
where $h^{*}$ is the same as in (189).
Instead of the symplectic form $\O_{j}$
on the cell $O^{j}_{h_{0}^{*}}$
we shall consider
its pull-back $p^{*}_{j}(h^{*}_{0})\O_{j}$
defined on $G_{j}(h^{*}_{0})$.
It is easy to obtain the answer,
using formula (112) for the symplectic form on $D_{ij}$.
We put the parameter
of the symplectic leaf
$g^{*}\! =\! g^{*}_{0}\! =\! const$.
It kills the second term and the rest
gives us the following answer:
\begin{equation}   p_{j}^{*}(h_{0}^{*})\O_{j}
=\frac{1}{2}<\o_{h^{*}}\!\stackrel{\wedge}{,}\o_{g}>
\ \  . \end{equation}
There is no manifest dependence on $d_{j}$ in (193),
but one must remember that $g$ takes values
in the very special subset of $G$ (190).
The dependence is hidden there.
Anyway, the final result of our investigation
is quite elegant.
Each orbit of the dressing transformations
in $F_{R}$ splits into the sum
of symplectic leaves (188)
and the symplectic form on each leaf
can be represented in the uniformed way (193).

As in section 3 one can check independently
that two-form (193) is really a pull-back
of some closed form on $O^{j}_{h_{0}^{*}}$.
We suggest this proposition as an exercise
for an interested reader.

We have classified symplectic leaves
in the quotient space $F_{R}=D\! /\! G$
and in particular in its maximum cell
$G_{0}^{*}=G^{*}\! /\! \cal E$.
In this content the idea
to find symplectic leaves in the group $G^{*}$ itself
arises naturally. To this end let us consider
the following sequence of projections
$G_{U}^{*}\rightarrow G^{*}\rightarrow G_{0}^{*}$,
where $G_{U}^{*}$ is a universal covering group
of the group $G^{*}$. The group $G_{U}^{*}$
is a Lie-Poisson group. The Poisson bracket
on the group $G_{U}^{*}$ is defined uniquely
by the Lie commutator in $\g$ \cite{6}.
The covering $G_{U}^{*}\rightarrow G_{0}^{*}$
appears to be a Poisson mapping. Using this property
one can check that $G^{*}$ is a Lie-Poisson group
and the corresponding Poisson bracket makes
both projections $G_{U}^{*}\rightarrow G^{*}$
and $G^{*}\rightarrow G_{0}^{*}$ Poisson mappings.
It implies that  symplectic leaves in $G_{U}^{*}$
and in $G^{*}$ are connected components of preimages
of symplectic leaves in $G_{0}^{*}$.
Corresponding symplectic forms can be obtained
by pull-back from (193). On the other hand,
the formula (193) gives an expression
for symplectic forms on the leaves in $G_{U}^{*}$
and $G^{*}$, if we treat $h^{*}$ as an element
of one of these groups and $g$ as an element of $G_{U}$,
universal covering group of $G$.
Then we define the action of $G_{U}^{*}$ on $G_{0}^{*}$
by the formula (189) ($g$ is a projection to $G$
of some element $g_{U}\in G_{U}$) and lift the action
of $G_{U}$ from $G_{0}^{*}$ to $G_{U}^{*}$ or $G^{*}$.
It is always possible by the definition
of the universal covering group.
We can identify symplectic leaves in $G^{*}_{U}$
or $G^{*}$ with orbits of the action of $G_{U}$,
which we have just defined.

It is remarkable that in the deformed case
the groups $G$ and $G^{*}$ may be considered
on the same footing. Formula (193) defines
symplectic structure on the orbit of $G^{*}$-action
in $D\! /\! G^{*}$ as well as on the orbit of $G$-action
in $D\! /\! G$. The only thing we have to change
is the relation between $g$ and $h^{*}$:
\begin{equation}  h^{*}g_{0}d_{i}G^{*}=gd_{i}G^{*}
\ \  . \end{equation}

To consider the classical limit
we can introduce a deformation parameter
into the formula (193):
\begin{equation}
p_{j}^{*}(h_{0}^{*})\O_{j}^{\gamma }
=\frac{1}{2\gamma }<\o_{h^{*}}\!\stackrel{\wedge}{,}\o_{g}>
\ \  .  \end{equation}
In this way one can recover
the classical Kirillov form (36)
as we did it for $T^{*}\! G$ in section 3.

\section{Examples.}

In this section we shall consider two concrete examples
to clarify  constructions described in sections 2--4.

{\bf 1.} The first example concerns the Borel subalgebra
$\b_{+}$ of semisimple Lie algebra $\g$.
The algebra $\b_{+}$ consists
of Cartan subalgebra $\h\!\subset\! \g$
and nilpotent subalgebra $\n_{+}$
generated by the Chevalley generators
corresponding to positive roots.
In the simplest case $\g=sl(n)\, \, $ $\b_{+}$
is just an algebra of traceless upper triangular matrices.
We may define the projection $p:\b_{+}\rightarrow \h$.
Let us call $p(\e)\in \h$ a diagonal part of $\e$
and denote it $\e_{d}$.

The dual space $\b_{+}^{*}$ can be identified
with another Borel subalgebra \mbox{$\b_{-}\!
\subset\! \g $}, where $\b_{-}\!\! =\! \h\! +\!
\n_{-}$ includes the nilpotent subalgebra $\n_{-}$
corresponding to negative roots. The canonical pairing
of $\b_{+}$ and $\b_{-}$ is given
by the Killing form $K(x,y)\equiv Tr(xy)$ on $\g$:
\begin{equation}  <x,\e >=K(x,\e )+K(x_{d},\e_{d})
\ \  . \end{equation}
The natural commutator on $\b_{+}^{*}=\b_{-}$
defines a structure of bialgebra on $\b_{+}$.
The double $\d$ is isomorphic to the direct sum
of $\g$ and $\h$:
\begin{equation} \d(\b_{+})\simeq \g \oplus \h
\ \  ,  \end{equation}
Isomorphism (197) looks like follows:
\begin{equation}
(x,\e )\longrightarrow (x+\e ,x_{d}-\e_{d})  \ \ .
\end{equation}
The first component of the r.h.s. in (198)
belongs to $\g$ and satisfies the corresponding
commutation relations, while the second component
is an element of $\h$. Elements of $\d$,
satisfying the conditions
\begin{equation}  x=x_{d}\ \ ,\ \ \e =\e_{d}
\ \ ,\ \ x_{d}+\e_{d}=0
\ \ ,  \end{equation}
belong to the center of $\d$.

The group $D$ in this case is a product
of semisimple Lie group $G$
and its Cartan subgroup $H$:
\begin{equation}  D=G\times H \ \  .
\end{equation}
The groups $B_{+}$ and $B_{-}$, corresponding
to the algebras $\b_{+}$ and $\b_{-}$,
can be embedded into $D$ as follows:
\begin{equation}
B_{+}\longrightarrow (B_{+},(B_{+})_{d})
\ \  , \end{equation}
\begin{equation}
B_{-}\longrightarrow (B_{-},(B_{-})_{d}^{-1})
\ \ ,\end{equation}
where $(B_{+})_{d}$, $\, (B_{-})_{d}$
are diagonal parts
of the matrices $B_{+}$, $\, B_{-}$.
The decomposition (73) in this case
may be described more precisely:
\begin{equation} D=\bigcup_{i\in W}B_{+}W_{i}B_{-}
\ \ ,  \end{equation}
where $W$ is Weyl group of $G$
and the pair $W_{i}=(w_{i},I)$
consists of the element $w_{i}$ from $W$
and the unit element $I$ in $H$.
For nontrivial $w_{i}$
spaces $V(W_{i})$, $\, V^{*}(W_{i})$ (95), (96)
are nonempty.

For the algebras $\b_{+}$ and $\b_{-}$
we can use matrix notations (18), (19)
for the Maurer-Cartan forms. For example,
\begin{equation}
\o_{B_{+}}=(dB_{+}B_{+}^{-1},db_{+}b_{+}^{-1})
\ \ ,  \end{equation}
\begin{equation}
\m_{B_{-}}=(B_{-}^{-1}dB_{-},-b_{-}^{-1}db_{-})
\ \  .  \end{equation}
Here $b_{+}$ and $b_{-}$
are diagonal parts of $B_{+}$
and $B_{-}$ correspondingly.
The invariant pairing $<\,,>_{\d}$
acquires the form:
\begin{equation} <(g_{1},h_{1}),(g_{2},h_{2})>_{\d}
=Tr(g_{1}g_{2}-h_{1}h_{2}) \ \  . \end{equation}
Now we can rewrite form (112) on the cell $D_{ij}$
in this particular case:
\begin{equation}
d=(B_{+}w_{i}B_{-},(B_{+})_{d}(B_{-})_{d}^{-1})
=(B_{-}^{'}w_{j}^{-1}B_{+}^{'}
,(B_{-}^{'})_{d}^{-1}(B_{+}^{'})_{d})   \end{equation}
\begin{eqnarray} p_{ij}^{*}\O_{ij}
=\frac{1}{2}Tr(dB_{-}^{'}B_{-}^{'-1}
\wedge dB_{+}B_{+}^{-1}+db_{-}^{'}b_{-}^{'-1}
\wedge db_{+}b_{+}^{-1}+\nonumber  \\
+B_{-}^{-1}dB_{-}\wedge B_{+}^{'-1}dB_{+}^{'}
+b_{-}^{-1}db_{-}\wedge b_{+}^{'-1}db_{+})
\ \ . \end{eqnarray}

We have the symplectic structure on $D_{+}$
and it is interesting to specialize Poisson bracket (79)
for this case. We use tensor notations
and write down the Poisson bracket
for matrix elements of $d$ and $h$, $(d,h)\in D$:
\begin{equation} \{ d^{1},d^{2}\} =-(r_{+}d^{1}d^{2}
+d^{1}d^{2}r_{-}) \ \ , \end{equation}
\begin{equation} \{ d^{1},h^{2}\} =-(\rho d^{1}h^{2}
+d^{1}h^{2}\rho)  \ \ , \end{equation}
\begin{equation} \{ h^{1},h^{2}\} =0
\ \  . \end{equation}
Here $r_{+}$ and $r_{-}$ are
the standard classical $r$-matrices,
corresponding to the Lie \mbox{algebra $\g$:}
\begin{equation} r_{+}=\frac{1}{2}\sum h_{i}
\otimes h^{i}+\sum_{\alpha \in \Delta_{+}}e_{\alpha }
\otimes e_{-\alpha } \ \  , \end{equation}
\begin{equation} r_{-}=-\frac{1}{2}\sum h_{i}
\otimes h^{i}-\sum_{\alpha \in \Delta_{+}}e_{-\alpha }
\otimes e_{\alpha }  \ \ , \end{equation}
and $\rho$ is the diagonal part of $r_{+}\,$ :
\begin{equation}
\rho =\frac{1}{2}\sum h_{i}\otimes h^{i}
\ \  .\end{equation}

As a result of general consideration
we have obtained the symplectic structure
corresponding to nontrivial Poisson bracket (209)--(211).
At this point we leave the first example and pass
to the next one.
\\

{\bf 2.} Now we take a semisimple Lie algebra $\g$
as an object of the deformation.
It is the most popular and interesting example.
The dual space $\g^{*}$ may be realized
as a subspace in $\b_{+}\!\oplus \b_{-}$:
\begin{equation} \g^{*}=\{ (x,y)\in \b_{+}\!\oplus \b_{-}
\ \ ,\ \ x_{d}+y_{d}=0\}  \ \ .  \end{equation}
The pairing between $\g$ and $\g^{*}$ is the following:
\begin{equation}  <(x,y),z>=Tr\{ (x-y)z\}
\end{equation}
and the Lie algebra structure on $\g^{*}$
is inherited from $\b_{+}\!\oplus \b_{-}$.
It is easy to prove that the algebra double
is isomorphic to the direct sum of two copies
of $\g$ \cite{6}:
\begin{equation} \d \simeq\g \oplus \g \end{equation}
\begin{equation} \{ x,(y,z)\} \longrightarrow(x+y,x+z)
=(d,d^{'}) \ \ ,  \end{equation}
\begin{equation}  <(d_{1},d_{1}^{'})
,(d_{2},d_{2}^{'})>_{\d}=Tr(d_{1}d_{2}
-d_{1}^{'}d_{2}^{'}) \ \  , \end{equation}
where $x\in \g\ $, $\ (y,z)\in \g^{*}$.
Therefore, the group double $D$
is a  product of two copies of $G$:
\begin{equation}  D=G\times G \ \ .\end{equation}
The subgroups $G$ and $G^{*}$ can be realized in $D$
as follows:
\begin{equation}  G=\{ (g,g)\in D\} \ \ ,
\end{equation}
\begin{equation}  G^{*}=\{ (L_{+},L_{-})\in D\ \
,\ \ (L_{+})_{d}(L_{-})_{d}=I\}\ \ .\end{equation}

Any pair $(X,Y)\in D$ can be decomposed
into the product of the elements from $G^{*}$
and $G$ by means of the same Weyl group $W$:
\begin{equation}  X=L_{+}w_{i}g\ \ , \end{equation}
\begin{equation}  Y=L_{-}g\ \ .  \end{equation}
Here $(L_{+},L_{-})\in G^{*}\ $,$\ g\in G$ and $w_{i}$
is an element of the Weyl group $W$.
So we have the following decomposition:
\begin{equation} D=\bigcup_{i\in W}G^{*}W_{i}G
\ \ ,  \end{equation}
where $W_{i}=(w_{i},I)$.

In this example we do not consider
the symplectic structure on $D_{+}$
and pass directly to the description of orbits.
The space $F_{R}=D\! /\! G$ can be decomposed
as in general case:
\begin{equation}
F_{R}=\bigcup_{i\in W}(G^{*}\!\!/T_{-i})
\ \ , \end{equation}
where $T_{-i}$ is the subgroup of $B_{+}$,
generated by the positive roots, which transform
into the negative ones by the element $w_{i}$
of the Weyl group:
\begin{equation}  T_{-i}=\{ t\in B_{+}\ \
,\ \ t_{d}=I\ \ ,\ \ w_{i}^{-1}tw_{i}\in B_{-}\}
\ \ . \end{equation}
The dressing transformations
act on the space $F_{R}$ as follows:
\begin{equation} gL_{+}w_{i}=L_{+}^{g}w_{i^{g}}h
\ \ ,  \end{equation}
\begin{equation} gL_{-}=L_{-}^{g}h
\ \ ,  \end{equation}
where $(L_{+}^{g},L_{-}^{g})$ is the result
of the dressing action $AD^{*}(g)$ and $i^{g}$
is the index of the cell, where it lies.
By the general theory the symplectic leaves in $F_{R}$
are intersections of the cells $(G^{*}\! /T_{-i})$
and the orbits of the dressing transformations.
The analogue (193) of the Kirillov two-form
can be rewritten in the following form:
\begin{equation}
p_{j}^{*}\O_{j}=\frac{1}{2}Tr(dL_{+}L_{+}^{-1}
-dL_{-}L_{-}^{-1})\wedge dgg^{-1}\ \ . \end{equation}
It is convenient to define the matrix
\begin{equation}  L=L_{+}w_{i}L_{-}^{-1}
\ \ .  \end{equation}
It transforms under the action
of the transformations (228), (229) in a simple way:
\begin{equation}
L^{g}=L_{+}^{g}w_{i^{g}}(L_{-}^{g})^{-1}=gLg^{-1}
\ \ .  \end{equation}
Being an element of $G$, the matrix $L$
defines a mapping from $F_{R}$ to $G$
by means of the formula (231).
On each orbit of the conjugations (232)
we can find a matrix $L$ of canonical form.
Let us denote it by $L_{0}$:
\begin{equation}
L^{g}=gL_{0}g^{-1}=L_{+}^{g}w_{i^{g}}(L_{-}^{g})^{-1}
\ \ . \end{equation}
Using two different parametrizations
of the same matrix $L$, we can rewrite (230):
\begin{equation}
p_{j}^{*}\O_{j}=\frac{1}{2}Tr\{ g^{-1}dgL_{0}
\wedge g^{-1}dgL_{0}^{-1}+L_{+}^{-1}dL_{+}w_{j}
\wedge L_{-}^{-1}dL_{-}w_{j}^{-1}\}
\ \ . \end{equation}

Formula (234) was obtained for $w_{i}=I$
in the paper \cite{8} as a by-product
of the investigations of WZ model.
The first term in (234) is rather universal.
It depends neither on the choice
of the Borel subalgebra in the definition
of the deformation nor on the cell of $F_{R}$.
On the contrary, the second term keeps the information
about the particular choice of $(B_{+},B_{-})$ pair
and it depends on the element $w_{i}$
of the Weyl group characterizing the cell of the orbit.

It is instructive
to write down the Poisson bracket
for the matrix elements of $L$.
Using the classical $r$-matrices $r_{+},r_{-}$
(212), (213) and tensor notations, we have \cite{5}:
\begin{equation}  \{ L^{1},L^{2}\} =r_{+}L^{1}L^{2}
+L^{1}L^{2}r_{-}-L^{1}r_{+}L^{2}-L^{2}r_{-}L^{1}
\ \ . \end{equation}

Let us remind that the same symplectic form (230)
corresponds to another Poisson structure
\begin{equation}  \{ g^{1},g^{2}\}
=r_{+}g^{1}g^{2}-g^{1}g^{2}r_{+}
=r_{-}g^{1}g^{2}-g^{1}g^{2}r_{-}
\ \ ,  \end{equation}
if instead of conditions (228), (229)
we impose the following set of constraints
on $L_{+}$, $\, L_{-}$ and $g$:
\begin{equation} L_{+}gw_{i}=g^{L}w_{i^{L}}L_{+}^{'}
\ \ ,  \end{equation}
\begin{equation} L_{-}g=g^{L}L_{-}^{'}
\ \ .  \end{equation}

\section{Discussion.}

In this section we formulate several problems
related to the symplectic structures described
in the paper. The first of them concerns
the quantum version of the presented formalism.
In the classical case the Kirillov symplectic form
appears in the content of the theory
of geometric quantization.
Roughly speaking, some coadjoint orbits
of the group $G$ equipped with the Kirillov form
correspond to irreducible representations
of the Lie algebra $\g$. The cotangent bundle $T^{*}\!G$
 with its canonical symplectic structure
corresponds to the regular representation of $\g$.
Actually, we may restrict ourselves to the latter case
because all the particular irreducible representations
can be obtained from the regular one
by means of the reduction procedure.
For Lie-Poisson groups the problem is not so simple
even for $D_{+}$. After the quantization
the Poisson algebra (80) becomes the quantum algebra
of functions on $D_{+}$. Its basic relations
can be written in the following form:
\begin{equation} d^{1}d^{2}=Rd^{2}d^{1}R^{*}
\ \ ,  \end{equation}
where we use tensor notations, $R$ and $R^{*}$
are quantum $R$-matrices corresponding
to the classical counterparts $r$ and $r^{*}$.
The result we expect as an outcome
of geometric quantization
is an irreducible representation of the algebra (239)
corresponding to a symplectic leaf in $D_{+}$.
It is easy to find such a representation
for the main cell $D_{00}=GG^{*}\cap G^{*}G$.
Algebra (239) $Funk_{q}(D_{+})$
acts in the space $Funk_{q}(G)$. It is an analogue
of the standard regular representation
in the space of functions on the group $G$.
 The algebra $Funk_{q}(G)$
is defined by the basic relations \cite{9}
\begin{equation} Rg^{1}g^{2}=g^{2}g^{1}R
\ \ . \end{equation}
On the cell $D_{00}$ we can decompose
the element $d$ as a product
\begin{equation} d=gh^{*}=g^{*}h  \end{equation}
of elements from $G$ and $G^{*}$. Matrix elements
of $G$ act on the space $Funk_{q}(G)$
by means of multiplication and matrix elements
 of $G^{*}$ generalize differential operators.
The regular representation in $Funk_{q}(G)$
was considered in \cite{10},
where the quantum analogue
of the Fourier transformation was constructed.

We expect that representations corresponding
to other symplectic leaves $D_{ij}$
can be found and presented in a similar form.
This would give a good basis
for the geometric quantization
in the direct meaning of the word,
i.e. establishing of the correspondence between the
orbits and the quantum group representations.
For $G\! =\! SU(n)$ this correspondence
has been described in paper \cite{11}
by means of quantization of orbits
of the dressing transformations.
It is a simple case because for $G\! =\! SU(n)$
$\; D\! =\! GG^{*}\!\! =\! G^{*}G$ and orbits
are symplectic leaves. It should be mentioned
that this correspondence appears in a natural way
in the course of investigations of the quantum groups
representation theory for the deformation parameter $q$
being a root of unity. If $q^{N}=1$,
there exists an irreducible representation
of the deformed universal enveloping algebra
$U_{q}(\g)$
corresponding to any orbit
of dressing transformations \cite{12}.

Another problem which we would like to mention
is a possible application of the machinery
of sections 3 and 4 to physics.
Having the closed form $\O$,
we can solve at least locally the equation
\begin{equation} d\alpha =\O  \ \ . \end{equation}
The one-form $\alpha$ may be treated
as a lagrangian of some mechanical system
so that the action looks like follows:
\begin{equation}  S_{0}=\int \alpha \ \ .
\end{equation}
If we add an appropriate hamiltonian $H$,
we get a system with the action
\begin{equation}  S=\int (\alpha -Hdt)  \ \ .
\end{equation}
Symplectic structures described in sections 3 and 4
provide a wide class of dynamical systems (244).
For the classical groups
one obtains many interesting examples in this way.
Among them one finds the WZNW model
and the gravitational WZ model \cite{13}.
Realizing the same idea for the Lie-Poisson case,
one can hope to construct integrable deformations
of these systems.

\section*{Acknowledgements.}

We are grateful to L.D.Faddeev,
A.G.Reiman and K.Gawedzki for stimulating discussions.
We would like to thank M.A.Semenov-Tian-Shansky
for the guidance in the theory of Lie-Poisson groups.
The work of A.A. was supported
by the joint program of CNRS (France)
and Steklov Mathematical Institute (Russia).
He thanks Prof. P.K.Mitter
for perfect conditions in Paris.
We are grateful to Prof. A.Niemi
for hospitality in Uppsala
where this work was completed.

\end{document}